\documentclass[pra,reprint,showpacs,showkeys,showpacs,times,superscriptaddress]{revtex4-1}
\usepackage{amsmath,amssymb,amsthm,times,graphics,graphicx,bm,subfigure}
\usepackage[colorlinks,linkcolor=red]{hyperref}

\newcommand{\be}{\begin{equation}}
\newcommand{\ee}{\end{equation}}
\newcommand{\ben}{\begin{eqnarray}}
\newcommand{\een}{\end{eqnarray}}
\newcommand{\bes}{\begin{subequations}}
\newcommand{\ees}{\end{subequations}}
\newcommand{\bF}{\begin{figure}}
\newcommand{\eF}{\end{figure}}
\newcommand{\dg}{\dagger}

\newcommand{\arxiv}[2][arxiv:]{\href{http://arxiv.org/abs/#1#2}{#1#2}}

\newcommand{\avg}[1]{\langle #1 \rangle}
\def\ket#1{ | #1 \rangle}
\def\bra#1{{\langle #1 |  }}
\def\tr{ {\rm{Tr }}\,}
\newcommand{\proj}[1]{\mbox{$|#1\rangle \!\langle #1 |$}}

\def\p{\eta_p}
\def\d{\eta_d}

\begin{document}
\title{Quantum metrology with imperfect states and detectors}
\author{Animesh Datta}
\email{animesh.datta@physics.ox.ac.uk}
\affiliation{Clarendon Laboratory, Department of Physics, University of Oxford, OX1 3PU, United Kingdom}

\author{Lijian Zhang}
\affiliation{Clarendon Laboratory, Department of Physics, University of Oxford, OX1 3PU, United Kingdom}

\author{Nicholas Thomas-Peter}
\affiliation{Clarendon Laboratory, Department of Physics, University of Oxford, OX1 3PU, United Kingdom}

\author{Uwe Dorner}
\affiliation{Centre for Quantum Technologies, National University of Singapore, 3 Science Drive 2, 117543 Singapore, Singapore}
\affiliation{Clarendon Laboratory, Department of Physics, University of Oxford, OX1 3PU, United Kingdom}

\author{Brian J. Smith}
\affiliation{Clarendon Laboratory, Department of Physics, University of Oxford, OX1 3PU, United Kingdom}

\author{Ian A. Walmsley}
\affiliation{Clarendon Laboratory, Department of Physics, University of Oxford, OX1 3PU, United Kingdom}

\date{\today}

\begin{abstract}

Quantum enhancements of precision in metrology can be compromised by system imperfections. These may be mitigated by appropriate optimization of the input state to render it robust, at the expense of making the state difficult to prepare. In this paper, we identify the major sources of imperfection an optical sensor: input state preparation inefficiency, sensor losses, and detector inefficiency. The second of these has received much attention; we show that it is the least damaging to surpassing the standard quantum limit in a optical interferometric sensor. Further, we show that photonic states that can be prepared in the laboratory using feasible resources allow a measurement strategy using photon-number-resolving detectors that not only attains the Heisenberg limit for phase estimation in the absence of losses, but also deliver close to the maximum possible precision in realistic scenarios including losses and inefficiencies. In particular, we give bounds for the trade off between the three sources of imperfection that will allow true quantum-enhanced optical metrology.

\end{abstract}


\maketitle

\section{Introduction}

Measurements can be made more precise by using sensor designs based on quantum mechanical rather that classical physical principles. The proximate cause of this enhanced precision is the reduced measurement noise enabled by quantum entanglement.  The realization of these advantages therefore hinges upon the preparation of particular nonclassical states that encode the sensor state parameter in such a manner as to allow its determination with precision beyond the standard quantum limit (SQL)~\cite{glm04}. Given a quantum state, the ultimate limit on the attainable precision is provided by the quantum Cram\'er-Rao bound (QCRB) via the quantum Fisher information (QFI)~\cite{bc94}. Early theoretical efforts in quantum metrology centered around designing quantum states that saturate this bound.

A paradigm for quantum enhanced measurement is optical interferometry, in which the phase difference between two field modes is to be estimated. For fixed photon numbers and no losses, quantum states minimizing the QCRB are the so-called $N00N$ states, consisting of a superposition of $N$ photons in one mode and none in the other ~\cite{lkd02,nagata07,afek10}. Unfortunately, $N00N$ states are exponentially more vulnerable to losses than classical states, and quickly lose their capacity for enhanced sensing. More recently, the effects of loss in an interferometer has been considered~\cite{ksd10,dorner09,lee09,kacprowicz10}. The optimal states for lossy phase estimation are, not surprisingly, dependent on the exact value of the loss in the interferometer. Consequently, no universal scheme for their preparation is possible.

%Additionally, this maximum precision assumes the ability to perform on the final quantum state certain optimal measurements, given by the eigenvectors of the so-called symmetric logarithmic derivative (SLD), which captures the differential changes in the state along a trajectory generated by the parameter. Such a measurement always exists~\cite{bc94}, but is in general prohibitively complex, because not only does it involve projections onto entangled states, but also depends on the loss in the interferometer.

However, losses in the interferometer are not the only imperfections to be dealt with. Preparation of the input state may be inefficient, delivering only an approximate version of the desired probe state. Also, the detectors may not be efficient, nor implement the requisite measurement strategy. In this Letter, we encompass imperfect input state preparation and sensor output measurement into our analysis. We show, surprisingly, that such imperfections are more detrimental to sensor performance than internal losses. However, we are able to identify a class of states that are close to optimally robust against such imperfections, and yet are feasibly constructed in the laboratory. This gives hope that the challenges of a palpably nonclassical sensor may be operated in realistic conditions. Since any implementation of quantum metrology will inevitably have all three imperfections, our results identify the range of imperfections and losses under which we can still demonstrate an objective advantage over classical phase estimation. They also pinpoint exactly the tradeoffs and bottlenecks in the path of demonstrating quantum enhanced metrology under realistic conditions. Our work addresses the fundamental gap between the principle and practice of quantum metrology. Furthermore, our work illustrates the usefulness of non-maximally entangled states.

Our scheme, shown in Fig.~(\ref{fig:setup}), starts with $N$ photons in each of two modes given by $\ket{\Psi}=\ket{N}\ket{N}$, which can be generated in a heralded manner with nonlinear processes such as parametric downconversion and photon-number-resolving detectors (PNRDs)~\cite{ntp10}, incident onto a $50$:$50$ beam splitter. The resulting state (See Eq. (\ref{eq:HBinMZ})), which we denote HB($N$), was proposed by Holland and Burnett~\cite{hb93}, has a photon number variance quadratic in $N$, and thus capable of attaining the Heisenberg limit for phase estimation~\cite{bc94}. They are more feasible in terms of laboratory resources than  $N00N$ and optimal states, yet their performance in not drastically diminished in the presence of losses~\cite{dbb02}. Recent work has demonstrated a scalable route to prepare highly pure HB($N$) states, relying on production of Fock states without complex linear-optical networks~\cite{ntp10}. In contrast, $N00N$ states require not only the generation of $N$ photons, but also a manipulation of these photons by means of a complex linear-optical network \cite{noon_linear}. The output of such a network is probabilistic since it relies on a particular detection (or nondetection) event of ancillary photons. This success probability usually decreases exponentially with increasing photon numbers.  Schemes that can, in principle, generate $N00N$ states with high success probability require either high nonlinearity~\cite{Kapale07} or actively controlled cavities~\cite{McCusker09}, which challenge current technology.  This decreasing probability of production necessitates post-selection on the outcomes to exhibit any perceived quantum enhancements.

\section{Holland-Burnett states}

\begin{figure}
\resizebox{7cm}{2.0cm}{\includegraphics{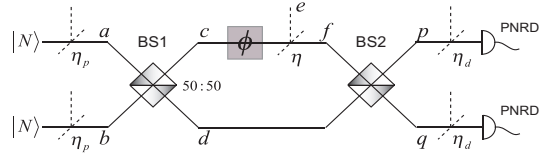}}
\caption{A schematic interferometer involving HB($N$) states. BS1 and BS2 are 50/50 beamsplitters, $\phi$ denotes the phase shift of mode $c$, and PNRD is a photon-number-resolving detector. $\eta$ is the loss in the interferometer arm, while $\eta_p$ and $\eta_d$ are the preparation and detection imperfections. $\eta = \eta_p = \eta_d =1$ denotes a perfect setup.}
\label{fig:setup}
\end{figure}

We show that for $HB(N)$ states, the QFI for phase estimation can be achieved with PNRDs. Fisher information also allows for an objective, situation-independent, resource-based certificate for our metrology scheme. We begin by calculating the QFI for phase estimation attainable with HB($N$) states in an ideal interferometer (Fig. (\ref{fig:setup})). After BS1, $\sqrt{2}a^{\dg} \rightarrow c^{\dg}+d^{\dg},\sqrt{2}b^{\dg} \rightarrow  c^{\dg}-d^{\dg}$, and the phase shifter $c^{\dg} \rightarrow e^{i\phi}c^{\dg}$,
\be
\label{eq:HBinMZ}
\ket{\Psi} =\sum_{n=0}^{N}A_n\ket{2n,2N-2n},A_n=\frac{\sqrt{2n!(2N-2n)!}}{2^N n!(N-n)!}e^{2in\phi},
\ee
where $\phi$ is the parameter to be estimated. The QFI quantifies changes in the initial state as a result of accumulating phase. This gives $ d\ket{\Psi}/d\phi \equiv  \ket{\Psi_{\phi}} = \sum_{n=0}^{N}2nA_n\ket{2n,2N-2n}$, leading to a QFI of~\cite{bc94}
\be
\label{eq:pureQFI}
\mathcal{J} = 4(\bra{\Psi_{\phi}}\Psi_{\phi}\rangle-|\bra{\Psi}\Psi_{\phi}\rangle|^2).
\ee
Since $\bra{\Psi_{\phi}}\Psi_{\phi}\rangle =N(3N+1)/2$, and $\bra{\Psi}\Psi_{\phi}\rangle = iN$,
\be
\label{eq:HBqfi}
\mathcal{J} = 2N(N+1).
\ee
This quantity, through the QCRB, $\Delta\phi \geq 1 / \sqrt{\mathcal{J}}$, provides the absolute attainable precision in phase estimation~\cite{bc94} using HB($N$) states. The quadratic behaviour of the QFI with the number of particles involved shows that we attain the Heisenberg limit. The original suggestion~\cite{hb93} of measuring the number difference in the two modes after BS2 (Fig. (\ref{fig:setup})) contains no information about the phase~\cite{cgb03}. A parity measurement $\Pi_N$ on one of the resulting modes leads to $\avg{\Pi_N}=P_N(\cos2\phi)$, where $P_N(\cdot)$ are Legendre polynomials. This provides a bound commensurate with Eq.~(\ref{eq:HBqfi}). Parity measurements are possible on the field mode~\cite{banaszek96}, but require additional resources including a local oscillator reference beam that is well matched to the probe state. Our endeavor here is to introduce a set of measurements that attains this limit, can be built from feasible laboratory resources, and is robust to imperfections such as inefficient detectors.

% The phase uncertainty is then given by $\Delta\phi=\sqrt{\avg{\Pi^2_N}-\avg{\Pi_N}^2}/|\partial \avg{\Pi_N}/\partial \phi| = 1/\sqrt{2N(N+1)}$. This

%\subsection{More practical measurements}
We show that a beam splitter and PNRDs suffices to saturate the QCRB.  Mixing modes $f$ and $d$ on BS2 yields  $\sqrt{2}f^{\dg} \rightarrow p^{\dg}+q^{\dg}, \sqrt{2}d^{\dg} \rightarrow p^{\dg}-q^{\dg}$.
%beyond which the state is
%$$
%\ket{\Psi}=\frac{1}{N!}\left(\frac{ie^{i\phi}\sin\phi}{2}\right)^{N}\!\!\!\!\!\!\left({p^{\dg}}^2 + {q^{\dg}}^2 -2ip^{\dg}q^{\dg}\cot\phi\right)^N \!\!\!\! \ket{00}.
%$$
%The expansion of the operators gives
%$$
%N!\left(\sum_{\mathcal{K}}\frac{(-2i\cot\phi)^{k_3}}{k_1!k_2!k_3!}{p^{\dg}}^{2k_1+k_3}\!\!\!{q^{\dg}}^{2k_2+k_3} \right) \ket{00},
%$$
%$\mathcal{K} = \{(k_1,k_2,k_3)|k_i \in \mathbb{Z}^+,k_1+k_2+k_3=N\}$.
Number resolving measurements $\ket{n}_p\ket{2N-n}_q$ on the two modes yields  $p_n = n!\left[P_N^{N-n}(\cos\phi)\right]^2/(2N-n)!,$ where $0\leq n \leq N$, and $P_N^l(\cdot)$ are the associated Legendre polynomials. The expression for $N \leq n \leq 2N$, is obtained by substituting $n \rightarrow 2N-n$. A simple yet interesting case is when we \emph{only} make the measurement $\ket{N}_p\ket{N}_q$. The resulting probability function $p_N =[P_N(\cos\phi)]^2$ has the same periodicity as the result of a parity measurement, and the Fisher information for this situation scales exactly as the Heisenberg limit in Eq.~(\ref{eq:HBqfi}), just like the parity measurement~\cite{slghog08}. Thus, the Heisenberg limit for phase estimation with lossless interferometers can be attained with just one pair of PNRDs. In essence, the external local oscillator necessary for the parity measurement has been replaced by the other arm of the interferometer, greatly simplifying the experimental demands. Photon number measurements still suffice when there are losses and imperfections, but the required number of measurements rises quadratically with $N$
%, and projection only onto $\ket{N}_p\ket{N}_q$ becomes suboptimal.

\section{Lossy interferometry}

\subsection{Loss in the interferometer}

\begin{figure}
\resizebox{8cm}{5.5cm}{\includegraphics{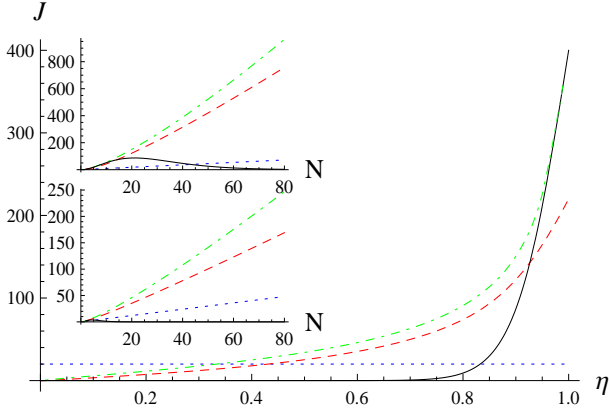}}
\caption{QFI for phase estimation as a function of the transmissivity $\eta$ for 20 input photons. Blue (Dotted): Standard quantum limit, Red (Dashed): HB(10) states, Black (Solid): $N00N$ states, Green (Dot-Dashed): Optimal states~\cite{dorner09}. Inset: QFI for phase estimation as a function of the photon number $N$ for $\eta = 0.9$ (top) , and $\eta = 0.6$ (bottom). }
\label{fig:HB5}
\end{figure}

Analysis of the performance of HB($N$) states in interferometry in the presence of losses starts with Eq.~(\ref{eq:HBinMZ}), the loss in a single arm of the interferometer being modeled as $c^{\dg} \rightarrow \sqrt{\eta}f^{\dg} + \sqrt{1-\eta}e^{\dg}$, $e$ being an unaccessible environment mode. Most of the loss occurs when the light interacts with the sample for phase accumulation, thus motivating treatment of loss in only one arm. Loss in both arms can be treated similarly, but requires numerical analysis and is beyond the scope of the current work. The subsequent state is
$$
\ket{\Psi} \!=\!\frac{1}{2^N}\sum_{n=0}^N\sum_{m=0}^{2n}C_nB_{n,m}\ket{2n-m}_f\ket{2N-2n}_d\ket{m}_e,
$$
where $C_n \!=\! 2n!\sqrt{(2N-2n)!}e^{2in\phi}/n!(N-n)!,$ $B_{n,m} =\eta^{n-m/2}(1-\eta)^{m/2}/\sqrt{(2n-m)!~m!},$
and $m$ is the number of photons lost to the environment. The resulting state obtained by tracing over mode $e$ is mixed, but its QFI can be calculated as shown in Sec. (\ref{sec:qfi}).
%Since this mode is to be traced over, we can rewrite the state as
%\be
%\label{eq:directsum}
%\ket{\Psi} = \sum_{m=0}^{2N}\ket{\psi_m}\ket{m}_e,
%\ee
%with $ \ket{\psi_m} =\frac{1}{2^N}\!\!\sum_{k=0}^{N-\lceil\frac{m}{2}\rceil}\!\!C_{k+\lceil\frac{m}{2}\rceil}B_{k+\lceil\frac{m}{2}\rceil,m}\ket{2k}_d\ket{2N-2k}_f,$ for $m$ even. For odd $m$, replace $2k\rightarrow 2k+1$ in the ket. Note that here and henceforth, we omit the explicit labeling of the modes for the sake of brevity, keeping in mind that they refer to modes $d$ and $f$. (See Fig.~(\ref{fig:setup})) Evaluation of the quantum Fisher information for phase estimation with the lossy states in Eq.~(\ref{eq:directsum}) is simplified by their block diagonal form. Setting $\ket{\tilde{\psi}_m} = \ket{\psi_m}/\sqrt{\mathfrak{N}_m}$, with $\mathfrak{N}_m=\bra{\psi_m}\psi_m\rangle$, we get
%$\mathcal{J}=\sum_{m=0}^{2N}\mathfrak{N}_m J(\ket{\tilde{\psi}_m}). $ Here $J$ is given by Eq.~(\ref{eq:pureQFI}), and leads to
%\ben
% J(\ket{\tilde{\psi}_m})&=& \frac{16}{2^{2N}\mathfrak{N}_m} \sum_{k=0}^{N-\lceil\frac{m}{2}\rceil}\left(k+\left\lceil\frac{m}{2}\right\rceil\right)^2C^2_{k+\lceil\frac{m}{2}\rceil}\times \nonumber \\
%&& B^2_{k+\lceil\frac{m}{2}\rceil,m}\left(1-\frac{C^2_{k+\lceil\frac{m}{2}\rceil}B^2_{k+\lceil\frac{m}{2}\rceil,m}}{\mathfrak{N}_m}\right).
%\een
%The summation can be performed in closed form, but the resulting expression in not very compact. We thus restrict our attention to some particularly interesting cases.
To start with, for $N=1$,
\be
\label{eq:HB1}
\mathcal{J}_{(N=1)} = 8\frac{\eta^2}{1+\eta^2},
\ee
which is the same as that obtained for two-photon $N00N$ states in~\cite{dorner09}, as expected, since they are identical to HB($1$) states. For higher photon numbers $N00N$ and HB($N$) states differ, and  HB($N$) states are more resilient to losses than the corresponding $N00N$ states with the same number of photons. This is shown in Fig.~(\ref{fig:HB5}) for $N=10,$ where the QFI for HB$(10)$ exceeds the standard quantum limit for $\eta > 0.45$ and adheres close to that of the optimal states for each value of the loss~\cite{dorner09}.

%\section{Nonideal preparation and detection}

\subsection{Imperfect state preparation}

We now analyse the performance of HB($N$) states in a more realistic situation where their preparation is not ideal. This is more than just with an eye towards experimental demonstration, though that provides part of the motivation.  More vital is our desire to address the gap between the theory and practice of quantum  metrology. To that end, we will work with the classical Fisher information $F$ obtainable with PNRDs. %This has led to several \textbf{slovenly} claims of quantum advantages in metrology.

We model a scenario where the input state might not necessarily be a twin Fock state $\ket{N}\ket{N},$ as in Fig. (\ref{fig:setup}). Independent of the physical nature of the probes, having exactly an equal number of bosons in two modes is difficult to realize experimentally.
%Recently, number uncertainties in atomic implementations of quantum metrology with HB($N$) states has been studied~\cite{mh09}. A number %difference of more than one atom was found to be severely detrimental to the quantum enhanced estimation of phase.
In an optical implementation, Fock states can be prepared by heralding~\cite{ntp10, mosley08}. In practice, the heralding efficiency is not unity. We can model this situation with ideal Fock state sources followed by a beam splitter of transmissivity $\eta_p$ in each mode before it is incident on the $50$:$50$ beam splitter. Such a beam splitter leads to
$
\ket{N} \rightarrow \rho\equiv \sum_{n=0}^N \left(
                                   \begin{array}{c}
                                     N \\
                                     n \\
                                   \end{array}
                                 \right)\p^n(1-\p)^{N-n} \proj{n}.
$
%The state after the $50$:$50$ beam splitter is $U(\rho_a\otimes \rho_b)U^{\dg}$, where $U = e^{i\pi(a^{\dg}b+ab^{\dg})/4}$. The phase accumulation operator is given by $P = e^{i\phi a^{\dg}a}$. PNRDs at the two output modes give $p_{mn} = \bra{m,n}U(P\otimes \mathbb{I}_b)U(\rho_a\otimes \rho_b)U^{\dg}(P\otimes \mathbb{I}_b)^{\dg} U^{\dg}\ket{m,n}.$  Also, $p_{mn}=0$ if $m+n >2N$.
The resulting classical Fisher information, assuming perfect transmission and detection with PNRDs, has a maximum for $\phi=0$, given by $F^{\max}_{\p} = 2N(N+1)\p^{N+1}.$ Interestingly, the minimum is attained for $\phi = \pi/2$, giving $ F^{\min}_{\p} = 2N(N+1)\p^{2N}.$
%The variation of the Fisher information as a function of $\phi$ is shown in Fig.~(\ref{fig:HB4}).

%\subsection{Imperfections all around}

\subsection{Imperfect detection}

Finally, we address the scenario where the detectors, PNRDs, are imperfect as well. This situation is modeled by placing beam splitters with transmissivity $\d$ in front of our PNRDs. We deal with the two simplest cases, $N=1$ and $N=2$ in order to illustrate key features of the system's performance. These will allow us to identify regimes within which we can unambiguously demonstrate quantum advantage in metrology, once again in a lossy scenario with nonideal sources and detectors. The procedure for obtaining the Fisher information for this scenario explicitly can be found in Sec. (\ref{sec:details}). When there is no loss in the interferometer, $F(\phi=0)=2N(N+1)(\p\d)^{N+1},$ illustrating the general principle that the quantum and classical Fisher information are symmetric under exchange of $\p$ and $\d$.

To judge the performance of HB($k$) state in providing genuine quantum advantage in phase estimation, we need to surpass the corresponding standard quantum limit, given by $F^{SQL} = 2k\eta\d.$ This is the standard quantum limit for a classical experiment performed on an apparatus identical to the quantum one, assuming that the classical (coherent) state can be prepared with certainty. The figure of merit for a quantum advantage is the ratio
\be
\label{eq:eth}
\eth_k(\p,\eta,\d) = \frac{F_{(N=k)}}{F^{SQL}} \geq 1.
\ee
We begin with HB($1$), in which case
\be
\label{ineq:HB1}
\eth_{1}(\p,\eta,\d) = \frac{4\p^2\d\eta}{1+\eta^2} > 1.
\ee
An expression such as this is very beneficial, as it demonstrates the tradeoffs involved in state preparation, interferometer construction, and detection imperfection, which allows experimentalists to direct their efforts appropriately. For instance, if $\d < 0.5$, its is impossible to beat the standard quantum limit with HB($1$) states, thereby rendering moot any discussion about the nature of the source and the interferometer. The asymmetry between preparation and detection imperfections in the final reckoning is due to the fact that the state attaining the SQL, a coherent state, can be produced with unit efficiency.

\begin{figure}
\resizebox{6cm}{6cm}{\includegraphics{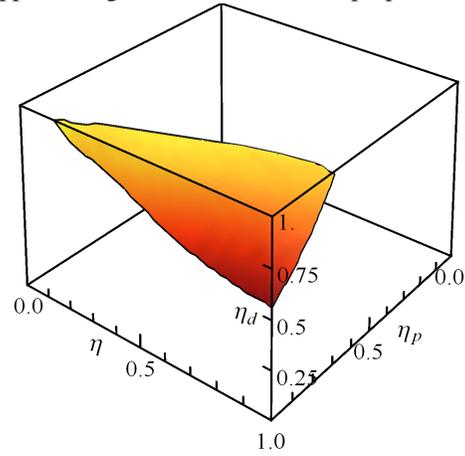}}
\caption{Plot of the feasibility region for beating the standard quantum limit using HB(2) states in the parameter space of preparation inefficiency, interferometer loss, and detector inefficiency $\p,\eta,\d$ respectively. The bottleneck in beating the standard quantum limit is the  detector imperfection, followed by the preparation imperfection and lastly, losses in the interferometer.}
\label{fig:FeasibleRegion2}
\end{figure}

The quantum advantage for HB($2$) is addressed by $\eth_{2}(\p,\eta,\d) = F_{(N=2)}/4\eta\d,$ where the right hand side is maximized over $\phi$. To get an idea of the requirements for an experiment, we find numerically that $0.687, 0.135$ and $0.547$ are the minimum values of $\p, \eta, \d$ respectively required to beat the SQL when the other two are unity. The complete region where $\eth_{2}(\p,\eta,\d) \geq 1$ is depicted in Fig.~(\ref{fig:FeasibleRegion2}). In general, higher photon number states are more resilient to losses in the interferometer but they also put stricter demands on $\p$ and $\d.$ Thus, with increasing photon  numbers, the feasibility region would shrink along the two axes denoting the imperfections, and extend along that denoting loss, as shown in Fig. (\ref{fig:FeasibleRegion13}). It is also easy to see that this particular pattern is universal. The detector and preparation imperfections are identical as far as $F_{(N=k)}$ is concerned, so we can think in terms of only $\p.$ As discussed previously, HB state are quite resilient to losses in the interferometer, but to achieve this performance relies on precisely preparing the twin-Fock state and performing Fock-state-projection measurements. Thus, $\p$, and consequently $\d$ has more stringent requirements than $\eta.$

%and $\eth_3(0.695,1,1) \approx \eth_3(1,0.089,1) \approx \eth_3(1,1,0.578) \approx 1$.

To experimentally realize an improvement over its classical counterpart, quantum phase estimation with HB states requires high-quality state preparation and detection in addition to low-loss interferometers. In a realistic experiment with $95$\% interferometer transmission, and $60$\% detection efficiency (at the high end for commercially available Silicon avalanche photodiodes), the HB($2$) state preparation must be better than $\p \geq 0.91$, which is well beyond the current state of the art~\cite{ntp10}. Using the best PNRDs available, with detection efficiencies approaching $0.98$~\cite{swn}, relaxes the preparation of the HB($2$) state to $\p \geq 0.71$, which is within the currently attainable values of $0.4  \leq \p \leq 0.85$~\cite{mosley08}.

\section{Conclusions}

We have identified benchmarks for the preparation, detection, and interferometer quality in a practical demonstration of quantum enhanced metrology. Most importantly, we have shown the first two of these to be the most detrimental to beating the SQL. We have shown that HB states deliver close to the best possible precision in the presence of all these imperfections and losses. Since a scalable route for preparation of the HB states has been proposed~\cite{ntp10}, we concluded that if one considers the whole gamut of issues involved in a metrological setup, including state preparation and the final measurement, and uses the objective tool of classical and quantum Fisher information, HB states and PNRDs provide a scalable and practically realizable setup for quantum enhanced metrology.

This work was funded in part by EPSRC (Grant EP/H03031X/1), the European Commission (FP7 Integrated Project Q-ESSENCE, grant 248095, and the EU-Mexico Cooperation project FONCICYT 94142) and the US European Office of Aerospace Research and Development (grant  093020).

\section{Appendix}

\begin{widetext}

\subsection{Quantum Fisher Information of a lossy $HB(N)$ state}
\label{sec:qfi}

Since mode $e$ of the state in Eq. (4) of the text is to be traced over, we can rewrite it as
\be
\label{eq:directsum}
\ket{\Psi} = \sum_{m=0}^{2N}\ket{\psi_m}\ket{m}_e,
\ee
with
\be
\ket{\psi_m} =\frac{1}{2^N}\!\!\sum_{k=0}^{N-\lceil\frac{m}{2}\rceil}\!\!C_{k+\lceil\frac{m}{2}\rceil}B_{k+\lceil\frac{m}{2}\rceil,m}\ket{2k}_d\ket{2N-2k}_f,
\ee
for $m$ even. For odd $m$, replace $2k\rightarrow 2k+1$ in the ket. The expressions for $B$ and $C$ are provided in the text. Evaluation of the quantum Fisher information for phase estimation with the lossy states in Eq.~(\ref{eq:directsum}) is simplified by their block diagonal form. Setting $\ket{\tilde{\psi}_m} = \ket{\psi_m}/\sqrt{\mathfrak{N}_m}$, with $\mathfrak{N}_m=\bra{\psi_m}\psi_m\rangle$, we get
$\mathcal{J}=\sum_{m=0}^{2N}\mathfrak{N}_m \mathcal{J}(\ket{\tilde{\psi}_m}). $ Here $J$ is given by Eq.~(2) in the text, and leads to
\be
\mathcal{J}(\ket{\tilde{\psi}_m})=
\frac{16}{2^{2N}\mathfrak{N}_m}\sum_{k=0}^{N-\lceil\frac{m}{2}\rceil}\left(k+\left\lceil\frac{m}{2}\right\rceil\right)^2C^2_{k+\lceil\frac{m}{2}\rceil} B^2_{k+\lceil\frac{m}{2}\rceil,m}\left(1-\frac{C^2_{k+\lceil\frac{m}{2}\rceil}B^2_{k+\lceil\frac{m}{2}\rceil,m}}{\mathfrak{N}_m}\right).
\ee

\subsection{Classical Fisher Information for lossy interferometer and imperfect sources and detectors}
\label{sec:details}

Let  $U_{ab}(\eta) = e^{i\theta(a^{\dg}b+ab^{\dg})}$ denote a beamplitter  across modes $a,b$ with transmissivity $\eta = \cos^2\theta.$ Then, using the abbreviation
\be
X\circ Y \equiv XYX^{\dg},
\ee
the state just after BS1 in modes $c$ and $d$,  $\sigma^1_{cd}$  is
\be
\sigma^1_{cd} = U_{ab}\circ (\rho_a\otimes \rho_b),
\ee
where $\rho_a = \sum_{n=0}^N \left(
                                   \begin{array}{c}
                                     N \\
                                     n \\
                                   \end{array}
                                 \right)\p^n(1-\p)^{N-n} \proj{n}.$
If $\vartheta_x$ denotes the vacuum in a mode $x,$ then the state after BS2 is given by
\be
\sigma^2_{pq}= \tr_{\!e}\!\left[U_{fd}\!\left(\frac{1}{2}\right)\circ U_{ce}(\eta) \circ(P_c\otimes \mathbb{I}_{de})\circ (\sigma_{cd}^1\otimes \vartheta_e)\right]
\ee
where $P_c = e^{i\phi c^{\dg}c}$ is the phase accumulation operator, and $\eta$ denotes the interferometer loss. If the vacuum modes associated with the lossy detectors on modes $p,q$ are labelled $p',q',$ then the probabilities at the two PNRDs are now given by
\be
p_{mn} = \bra{m,n}\tr_{p'q'}\!\left[(U_{pp'}(\d)\otimes U_{qq'}(\d))\circ(\sigma^2_{pq}\otimes\vartheta_{p'}\otimes\vartheta_{q'})\right]\ket{m,n},
\ee
where $m,n \geq 0$ and $m+n \leq 2N$. Additionally, $p_{mn} = p_{nm}$ Thus, there are in general $(N+1)^2$ independent measurement outcomes. The resulting classical Fisher information expressed as
\be
\label{eq:fish}
F = \sum_{m,n}\frac{(\partial p_{mn}/\partial \phi)^2}{p_{mn}},
\ee
is, in general, a function of the phase to be estimated $\phi.$ For $N=1,$ the probabilities of the different outcomes can be arranged in a matrix given by
\be
P_1 = \left(
        \begin{array}{ccc}
          p_{00} & p_{01} & p_{02} \\
          p_{10} & p_{11} & 0 \\
          p_{20} & 0 & 0 \\
        \end{array}
      \right),
\ee
where
\bes
\ben
p_{00} &=& 1 -(1+\eta)\p\d + \frac{1+\eta^2}{2}\p^2\d^2, \\
p_{01} &=&  \frac{1+\eta}{2}\p\d - \frac{1+\eta^2}{2}\p^2\d^2,\\
p_{02} &=&  \frac{1+\eta^2-2\eta\cos2\phi}{8}\p^2\d^2, \\
p_{11} &=&   \frac{1+\eta^2+2\eta\cos2\phi}{4}\p^2\d^2.
\een
\ees

\begin{figure}[h]
\resizebox{7cm}{7cm}{\includegraphics{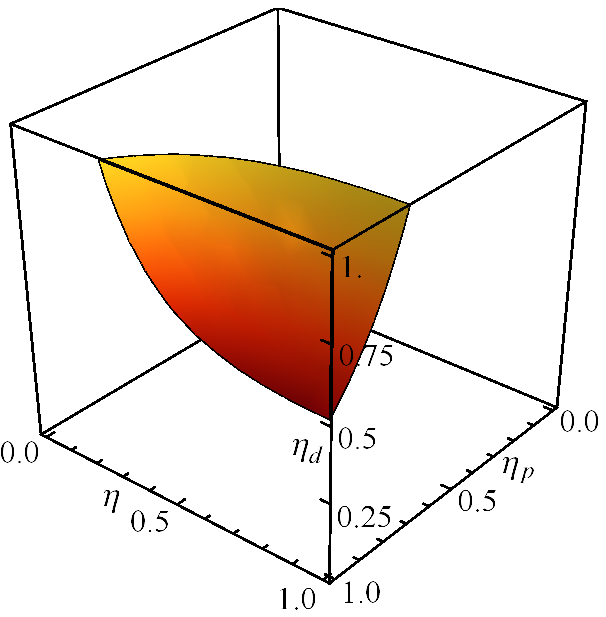}}
\hfill
\resizebox{7cm}{7cm}{\includegraphics{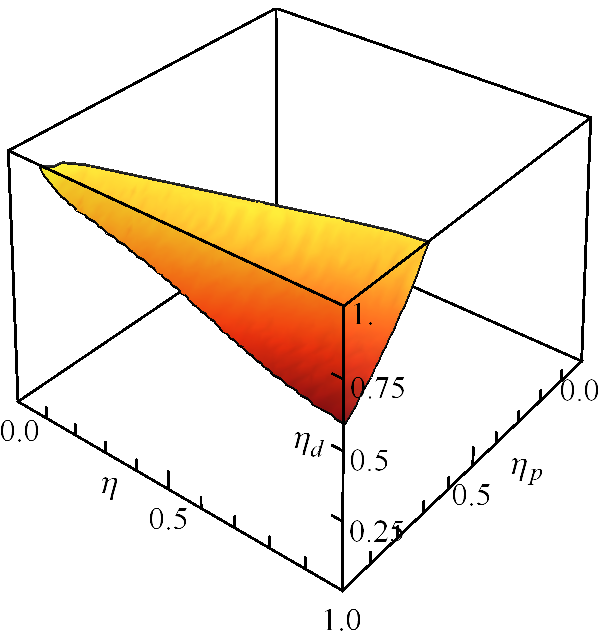}}
\caption{Plot of the feasibility region for beating the standard quantum limit using $HB(1)$ states [Left] in the parameter space of preparation inefficiency, interferometer loss, and detector inefficiency $\p,\eta,\d$ respectively. The bottleneck in beating the standard quantum limit is the  detector imperfection, followed by the preparation imperfection and lastly, losses in the interferometer. Same for $HB(3)$ states [Right].}
\label{fig:FeasibleRegion13}
\end{figure}

The classical Fisher information can easily be calculated using Eq. (\ref{eq:fish}), resulting in
\be
F_{(N=1)} = \frac{8\p^2\d^2\eta^2(1+\eta^2)\sin^22\phi}{1+\eta^4-2\eta^2\cos4\phi}.
\ee
This function is maximized at $\phi = \pi/4$ leading to
\be
F_{(N=1)} = \frac{8\p^2\d^2\eta^2}{1+\eta^2}.
\ee

For $N=2,$ the probabilities for the different measurement outcomes are
\be
P_2 = \left(
            \begin{array}{ccccc}
                 p_{00} & p_{01} & p_{02} & p_{03} & p_{04} \\
                 p_{10} & p_{11} & p_{12} & p_{13} & 0 \\
                 p_{20} & p_{21} & p_{22} & 0 & 0 \\
                 p_{30} & p_{31} & 0 & 0 & 0 \\
                 p_{40} & 0 & 0 & 0 & 0 \\
               \end{array}
             \right),
\ee
where
\bes
\ben
p_{00} &=& 1 -2(1+\eta)\p\d + \frac{5+2\eta+5\eta^2}{2}\p^2\d^2 - \frac{3+\eta+\eta^2+3\eta^3}{2}\p^3\d^3 + \frac{3+3\eta^2+2\eta^4}{8}\p^4\d^4, \\
p_{01} &=& (1+\eta)\p\d - \frac{5+2\eta+5\eta^2}{2}\p^2\d^2 + \frac{3+\eta+\eta^2+3\eta^3}{4}\p^3\d^3 - \frac{3+3\eta^2+2\eta^4}{4}\p^4\d^4, \\
p_{02} &=&  \frac{5+(4-6\cos2\phi)\eta+5\eta^2}{8}\p^2\d^2 - \frac{9+(5-6\cos2\phi)\eta(1+\eta)+9\eta^3}{8}\p^3\d^3  \nonumber\\
     && - \frac{9+10\eta^2+9\eta^4 - 6\eta(1+\eta^2)\cos2\phi}{16}\p^4\d^4, \\
p_{03} &=&  \frac{3(1+\eta)(1+\eta^2-2\eta\cos2\phi)}{16}\p^3\d^3 - \frac{3(1+\eta^2)(1+\eta^2-2\eta\cos2\phi)}{16}\p^4\d^4,\\
p_{04} &=&  \frac{3}{128}(1+\eta)(1+\eta^2-2\eta\cos2\phi)^2\p^4\d^4,\\
p_{11} &=&  \frac{5+6\cos2\phi\eta+5\eta^2}{8}\p^2\d^2 - \frac{9+(1+6\cos2\phi)\eta(1+\eta)+9\eta^3}{8}\p^3\d^3  \nonumber\\
     && - \frac{9+2\eta^2+9\eta^4 + 6\eta(1+\eta^2)\cos2\phi}{16}\p^4\d^4, \\
p_{12} &=& \frac{9+\eta+\eta^2+9\eta^3 +6\eta(1+\eta)\cos2\phi}{16}\p^3\d^3 + \frac{9+2\eta^2 +9\eta^4+6\eta(1+\eta^2)\cos2\phi}{16}\p^4\d^4,\\
p_{13} &=&  \frac{3}{32}(1+\eta^4-2\eta\cos2\phi)\p^4\d^4,\\
p_{22} &=&  \frac{1}{64}(9 + 4 \eta^2 + 9\eta^4 + 12(\eta + \eta^3) \cos2\phi + 18 \eta^2 \cos4\phi)\p^4\d^4.
\een
\ees
The classical Fisher information $F_{N=2}$ can once again be calculated using Eq. (\ref{eq:fish}). This expression is then used in the plotting of Fig. (\ref{fig:FeasibleRegion2}) in the text.

\end{widetext}

\end{document}